# Telecom-Integrated Photonic Memory Operating Near the Mechanical Ground State

Jindao Tang,[1] Bo Jing,[2,*] Liping Zeng,[1] Peiqin Chen,[1] Hengrui Liang,[1] Yifei Zhang,[1] Xinyao Xu,[1] Qizhi Cai,[1] Xueying Zhang,[3] Qiang Zhou,[1] You Wang,[4] Haizhi Song,[4] Guangcan Guo,[5,6] and Guangwei Deng[1,5,6,7,8,†].

[1] Institute of Fundamental and Frontier Sciences, University of Electronic Science and Technology of China, 610054 Chengdu, China
[2] Center for Information Photonics and Communications, School of Information Science and Technology, Southwest Jiaotong University, 611756 Chengdu, China
[3] School of Science, Key Laboratory of High Performance Scientific Computation, Xihua University, 610039 Chengdu, China
[4] Southwest Institute of Technical Physics, 610041 Chengdu, China
[5] Key Laboratory of Quantum Information, University of Science and Technology of China, 230026 Hefei, China
[6] Hefei National Laboratory, 230088 Hefei, China
[7] Key Laboratory of Quantum Physics and Photonic Quantum Information, Ministry of Education, University of Electronic Science and Technology of China, 610054 Chengdu, China
[8] Institute of Electronics and Information Industry Technology of Kash, 844000 Kash, China

**ABSTRACT**. Scalable quantum networks require quantum memories that are chip-integrated, telecom-band compatible, and capable of flexible retrieval. Nanofabricated mechanical resonators meet these criteria. They offer independent tunability of optical and mechanical modes, long-lived phonon states, and design flexibility beyond atomic systems, making them strong candidates for practical integrated quantum memory. Here, we demonstrate an on-chip, absorptive optomechanical memory for telecom-band photons, based on optomechanically induced transparency (OMIT) and operating near the mechanical ground state. The device stores telecom-band photons, demonstrating compatibility with external photon sources at the few-photon level, while enabling on-demand retrieval. By placing the device in a dilution refrigerator at 20 mK and tailoring the control field to suppress optical heating, we achieve a remarkably low phonon occupancy of just 0.32 during the storage process. Our results lay the groundwork for scalable, phonon-based quantum memory devices and open new avenues for integrating mechanical systems into practical quantum network architectures.

## I. INTRODUCTION.

The construction of scalable quantum networks [1-3] is a central goal in advancing quantum communication [4] and distributed quantum computing [5]. Such networks inevitably rely on the integration of diverse quantum devices — such as single-photon sources [6] and quantum memories [7,8] — to harness the unique strengths of different platforms. Among these essential building blocks, quantum memories play a pivotal role by enabling the synchronization of probabilistic events, facilitating entanglement distribution over long distances, and supporting the implementation of quantum repeaters [2,9,10]. To meet the stringent requirements of network-scale operation, quantum memories need to satisfy several critical criteria including operation at telecom wavelengths for minimal fiber transmission loss, integration capability for scalability, and efficient on-demand retrieval to coordinate photonic qubits across the network [11].

Despite extensive progress, existing quantum memory implementations still face significant limitations. Bulk cold atomic memories and trapped ion systems have already enabled node-to-node networking but rely on large free-space optical setups, which constrain integration and scalability[12-15]. To overcome these limitations, considerable efforts have been made toward developing integrated quantum memories using solid-state platforms. For example, color-center-based systems allow for chip-level integration and have shown promising performance, but their operation wavelengths are typically far from the telecom band, making them have to rely on frequency conversion for long-distance quantum communication[16]. In contrast, only rare-earth-ion-doped solids—particularly those doped with erbium—naturally operate at telecom wavelengths and are compatible with integrated photonic architectures[17-19]. Nevertheless, although these systems benefit from inherent telecom-band operation and compactness, the AFC mechanism for Kramers ions intrinsically lacks true on-demand retrieval capability, as the retrieval time is determined by the preprogrammed comb spacing and auxiliary energy levels required for spinwave-AFC are not available[20]. Attempts to realize on-demand retrieval through electric-field-induced Stark shifts offer limited tunability, severely constraining the flexibility of storage operations[21]. These challenges highlight the urgent need for alternative platforms that can simultaneously achieve integration, telecom compatibility, and flexible on-demand storage — key attributes for building practical quantum networks.

*Contact author: goubo@swjtu.edu.cn
†Contact author: gwdeng@uestc.edu.cn

Nano-optomechanical systems [22], such as optomechanical crystal (OMC) cavities [23-33], have recently emerged as promising candidates to address these challenges. These systems not only provide design flexibility for custom frequency and structural parameters, unconstrained by natural atomic resonances [34], but also enable thermal noise suppression through high-frequency mechanical modes[35,36]. Featuring strong optomechanical coupling, chip-scale fabrication compatibility, and telecom-wavelength operation, these platforms offer a distinctive route toward integrated quantum memories. Currently, room-temperature demonstrations of coherent light storage in optomechanical systems [37,38] remain fundamentally limited by significant thermal noise, which severely obscures quantum properties of nanomechanical resonators. Recent experiments utilizing membranes with ultra-high-quality-factor optomechanical resonators have shown improved efficiency and prolonged storage times [39]. However, the thermal noise level in this work remains firmly in the classical regime, and genuine quantum memory, particularly at the single-photon level, requires further cooling to suppress thermal excitation, and achieve thermal occupancy below unity. These limitations underscore the necessity of cryogenic environments for realizing low-noise, high-fidelity optomechanical quantum memories. While previous demonstrations had implemented DLCZ-type [40] emissive quantum memory in such systems under cryogenic conditions[41], this protocol generates nonclassical states internally and is not inherently designed for storing externally generated photonic qubits, making it difficult to fully harness the advantages of high-quality external single-photon sources [6,42-45]. Consequently, exploiting cryogenic mechanical resonators to directly store external telecom-band photons via an absorptive protocol like Optomechanically Induced Transparency (OMIT)—while simultaneously enabling on-demand retrieval and chip-scale integration—presents a compelling approach for advancing quantum networks based on hybrid quantum devices. Yet, this remains an unachieved milestone.

In this work, we place an integrated nano-mechanical resonator in a dilution refrigerator at 20 mK and demonstrate few-photon memory with on-demand retrieval using the OMIT method. Via manipulating the control light, a coherent signal photon pulse at the few-photon level (41 photons per pulse) is stored in the nano-device and retrieved on demand. By increasing the pulse duration and reducing the intensity of the control light, while minimizing heating effects, we maintain a certain level of storage efficiency, thereby enabling the observation of the storage phenomenon. The internal memory efficiency is measured at 4.2%, while this efficiency near the quantum ground state is $5\times10^{-5}$. The storage time here is 150 ns and can be further extended by increasing acoustic confinement. Crucially, the mechanical resonator operates near the quantum ground state with sub-unity mean thermal occupancy ($\langle n \rangle < 1$), effectively suppressing thermal noise that obscures quantum behavior. Our results mark a significant step toward integrating mechanical memories into quantum information architectures.

## II. METHODS

### A. The principle of OMIT memory

Optomechanically Induced Transparency is a light-induced transparency phenomenon that occurs in mechanical resonators[25,46], analogous to electromagnetically induced transparency (EIT) in atomic systems. The central idea of the present work is to replace the atomic lambda system for EIT memory with optomechanical resonators and to exploit the long coherence times [41] induced by the mechanical modes. The optomechanical system adopted here is a one-dimensional optomechanical crystal cavity. Figure 1(a) shows the finite element simulation diagrams of the optical and mechanical modes of concern in the experiment. The interaction between coherent phonons and optical fields can also be utilized for photon storage and can be represented as an effective three-level system in terms of energy levels, as illustrated in Fig. 1(b). The energy transition between |n+1, 0⟩ and |n, 0⟩ corresponds to the signal photons (represented by the yellow arrow), while the transition between |n+1,0⟩ and |n,1⟩ corresponds to the control light (represented by the red arrow). When the control field is applied, the OMIT effect renders the medium transparent to the signal photons. If the signal photons are simultaneously applied and the control field is adiabatically turned off, the signal photons can be coherently stored in the optomechanical system. Upon reintroducing the control light, the stored photons can be retrieved.

*Contact author: goubo@swjtu.edu.cn
†Contact author: gwdeng@uestc.edu.cn

Figures 1(c) and 1(d) depict the frequency-domain and time-domain descriptions of the memory, respectively. In the frequency domain, the control field is detuned to the red side of the optical cavity, while the signal field's frequency is centered within the cavity. The detuning between the two is on the order of the mechanical frequency. During the time-domain process, a strong optical pulse (control field) is injected to the system while simultaneously superimposing the signal field. After a delay time of $t_i$, the control light is reapplied. The stored mechanical excitation is then converted into photons, known as the retrieval process. The retrieved signal photons co-propagate with the control field, and this noise issue can be addressed via filtering.

### *2. OMIT in OMC*

We first employed continuous-wave measurements to characterize the OMIT phenomenon. The experimental setup is shown in Fig. 2(a). The device under test consists of a one-dimensional optomechanical crystal microcavity fabricated from the silicon device layer of a silicon-on-insulator (SOI) wafer. The optical cavity exhibits a resonance at 1538 nm, while the mechanical mode resonates at 5314.1 MHz (See Supplemental Material, Sec. 10 [46]). The device was housed within a dilution refrigerator maintained at 20 mK, and was accessed optically via on-chip waveguides. As depicted in Fig. 2(a), a laser stabilized to the red sideband of the optical mode served as the control field. The laser was split via a beam splitter for power monitoring, and modulated by an electro-optic modulator (EOM) to generate sidebands at tunable frequency offsets. The modulation signal was provided by one port of a vector network analyzer (VNA). While two sideband photons were symmetrically distributed, one of the sidebands lay far outside the cavity linewidth and was thus neglected. The remaining sideband, near-resonant with the cavity, coupled to the mechanical mode. After interacting with the microcavity, the remaining signal photons, along with the control light, were fed back to a high-speed photodetector for heterodyne detection. As the control light maintained constant amplitude and frequency, variations in the reflected signal amplitude—arising from its frequency shift due to optomechanical interaction—were converted into electrical signals and analyzed by the VNA.

The VNA was configured to measure the $S_{21}$ parameter, which represents the ratio of the intensity of the radio frequency signal converted from the optical modulated sideband passing through the optomechanical system to the intensity of the original radio frequency modulated signal. By scanning the modulation frequency, the test quantified the frequency-dependent transmission response of the device. As shown in Fig. 2(b), distinct dips in the transmission correspond to optical resonances. When the two-photon resonance condition ($\Delta\omega = \omega_m$) is satisfied, destructive interference suppresses the intracavity buildup of the signal field. This interference does not rely on resonance between natural energy levels and can thus be extended to previously inaccessible wavelength regions by designing the optical mode. The simultaneous presence of the control and signal photons induces radiation pressure oscillations with a frequency of $\Delta\omega$. When the oscillation frequency of this driving force approaches the mechanical resonance frequency ($\omega_m$), the mechanical mode will begin to oscillate coherently. These coherent oscillations lead to optomechanical scattering of the control light field within the cavity, and the re-modulation of the control light field also remains coherent. If the system is in the resolved sideband regime, Stokes scattering of the control light is suppressed. In this case, only anti-Stokes scattering within the cavity needs to be considered. When the anti-Stokes scattered photons become spectrally degenerate with the signal photons, destructive interference will give rise to an induced transparency window. As shown in Fig. 2(b) and 2(c), a peak can be observed at the bottom of the optical cavity dip. The frequency and linewidth of this peak directly reflect the mechanical resonator's properties within the optomechanical system.

As shown in Fig. 2(b), two transparency windows were observed within the dip corresponding to the first- and second-order breathing modes of the mechanical resonator. The mode indicated by the red arrow coincided with the frequency measured independently by spectrum analysis. The transparency window shown in Fig. 2(c) was obtained at a control light power of -15 dBm, with a linewidth of 332 kHz. Further experiments investigated the relationship between control light power, transparency window width, and depth. The OMIT depth ( $d$ ) can be defined as the normalized power transmission on two-photon resonance in the optical resonance dip:

$$d = (r_{omit}^2 - r_{op}^2)/(1 - r_{op}^2) \tag{1}$$

where $r$ is reflection coefficient of the detected field amplitude, calculated as follows.

*Contact author: goubo@swjtu.edu.cn
†Contact author: gwdeng@uestc.edu.cn

$$r = 1 - \frac{\kappa_{ex}}{i(\Delta - \omega_S) + \kappa/2 + \dfrac{(g_0\sqrt{\bar{n}_{cav}})^2}{i(\omega_m - \omega_S) + \gamma_i/2}} \tag{2}$$

$n_c$ is the photon number in optical cavity, determined by control light intensity. The optical mode is characterized by a total damping rate $\kappa = \kappa_i + \kappa_{ex}$. $\Delta$ is the pump detuning from the optical cavity. Here, $r_{op}$ is the reflected amplitude of optical resonance ( $\Delta - \omega_S$ ) ignoring the mechanical response at that time. $r_{omit}$ is the reflected amplitude when both resonance $\Delta - \omega_S = 0$ and $\omega_m - \omega_S = 0$ are satisfied. As shown in Fig. 2(d), at a control light power of -10 dBm, the transparency window exhibited a depth of 97% (approaching complete transparency) with a linewidth of 877 kHz. These results indicated that increased optical power expands the bandwidth and depth of the transparency window. When the laser power was reduced to -30 dBm, the window depth decreased to 12% with a corresponding linewidth of 37 kHz.

For photon storage applications utilizing our one-dimensional optomechanical crystal microcavities, the intense control light required for OMIT can readily induce thermal heating, thereby overwhelming the mechanical quantum state with thermal noise and degrading system performance[24,35,36]. While pulse shortening can mitigate heating effects, the consequent spectral broadening compromises bandwidth matching and photon coherence (See Supplemental Material, Sec. 1 [46]). Based on the above results, the depth of the OMIT window does not decrease linearly with the control field intensity[25,47]. Given that the optical cavity operates near the critical coupling regime ( $\kappa_i = \kappa_{ex}$ ), the depth of the OMIT window can be expressed as: $(4\bar{n}_{cav}g_0^2)^2/(4\bar{n}_{cav}g_0^2 + \Gamma_m\kappa)^2$ . This expression reveals a positive correlation between window depth and photon number. When $\bar{n}_{cav} \sim \Gamma_m\kappa/4g_0^2$, the depth increases rapidly with the increase in the number of photons. However, as $\bar{n}_{cav}$ further increases, this growth tends to slow down. When $\bar{n}_{cav} \gg \Gamma_m\kappa/4g_0^2$ , the window depth approaches 100%. The theoretical analysis shows that there is an optimal range of control light intensity in which substantial OMIT depth is maintained, even at minimal intensity levels (See Supplemental Material, Sec. 2 [46]). Under conditions of bandwidth matching, the reduction of the control light pulse intensity and appropriate extension of its duration can mitigate heating effects, keeping the system within the quantum ground state while preserving photon storage capability.

Although reducing control light power mitigates thermal effects, this approach inevitably compromises storage efficiency due to a decrease in the depth of the transparency window. Moreover, the concomitant narrowing of spectral bandwidth imposes stricter temporal requirements, as pulse duration must increase proportionally to maintain spectral overlap - a direct consequence of Fourier-transform limitations. As shown in Fig. 2(e), even with a 100-fold reduction in control light power, the depth of the transparency window remains at 12%. By appropriately increasing the pulse duration, a trade-off between maintaining storage efficiency and minimizing optical heating can be achieved (See Supplemental Material, Sec. 3 [46]). Therefore, by simultaneously lowering light intensity and extending pulse duration, the total heating power can be effectively reduced, potentially suppressing thermal noise to levels approaching the quantum ground state.

Since laser illumination can induce heating in the sample, as discussed in previous studies [35], a pulse-based approach was employed in subsequent experiments to suppress thermal noise while preserving signal integrity [24,36]. As shown in Fig. 3(a), the laser, locked to the red-detuned sideband of the optical cavity, passed through an acousto-optic modulator (AOM) to generate two square wave pulses during each experimental cycle. The first pulse (write pulse) was further modulated by an EOM to generate sidebands, which carried the signal photons. These signal photons were resonant with the center of the cavity mode, and thus lay within the center of the OMIT window. The second pulse (read pulse) remained unmodulated. The light pulse reflected from the sample were filtered to remove residual control light before detection. To ensure complete thermal dissipation between pulses at a high repetition rate, all pulsed storage experiments were conducted using a specific device with a deliberately tailored moderate phonon lifetime (See Supplemental Material, Sec. 4). All pulse sequences were governed by a programmable pulse generator, synchronized with the single-photon signal acquisition. To achieve sufficient suppression of background light, four cascaded free-space Fabry–Perot cavities were used, providing an extinction ratio exceeding 90 dB. The signal recorded by the single-photon detector included both the signal photons generated during the write pulse and the retrieved photons from the read pulse, which were temporally separated. A schematic of the complete optical path is provided in Supplemental Material, Sec. 5 [46].

*Contact author: goubo@swjtu.edu.cn
†Contact author: gwdeng@uestc.edu.cn

Figure 3(b) shows the spectral distribution of the write pulses, read pulses and the filter. The control light was effectively suppressed before reaching the single-photon detector, allowing both the input signal photons and the retrieved photons to be detected. Figure 3(c) illustrates the photon storage dynamics. Here, three different control light intensities were applied, which will result in varying retrieved signal amounts. During the storage process, a write pulse with a duration of 2500 ns and a read pulse with a duration of 1050 ns were sequentially injected into the sample. This specific delay was chosen to separate the write and read pulses in the time domain—eliminating optical cross-talk while maintaining a strong retrieved signal within mechanical lifetime. Crucially, our protocol enables continuous on-demand retrieval, meaning the read pulse can be triggered at any arbitrary moment after writing, which is inherently corroborated by our lifetime measurements. The only restriction is the natural exponential decay of the stored phonon state.

The write pulse contained both control light and signal photons; the control light structured an OMIT window, enabling the storage and subsequent retrieval of the signal photons. The signal photons were generated by modulating the control light using an EOM. During the input stage, the detected photon count exhibited a gradual decline, which arose from the deadtime (~100 ns) of the single-photon detector. At the onset of the pulse, the detector operated at full efficiency. However, as the pulse continued, the likelihood of missing photons increased due to the detector's intrinsic dead time. Notably, both the input and retrieved signal photons propagated through the same optical path and experienced identical losses. Consequently, the photon counts measured at the single-photon detector could be reliably used to determine the storage efficiency. To isolate the contribution of the memory process, a reference measurement was performed by injecting a pulse containing only control light. By subtracting the detection signal obtained without signal photons from that with signal photons, the contribution from memory process could be isolated, effectively revealing the retrieved signal photons. Here, the heating effect caused by the signal photons was negligible (See Supplemental Material, Sec. 6 [46]). During the retrieval phase, the detection rate of photons gradually decreased over time, indicating the progressive retrieval of the stored signal. By the end of the pulse, the count rate dropped by nearly an order of magnitude, suggesting that the majority of the stored photons had been successfully retrieved. Based on a comparison with the input signal photon number, the memory efficiency of our integrated device was determined to be 4.2%.

## III. RESULTS

In the photon storage dynamics experiment, high-energy (peak power = -5.7 dBm) and long-duration (1050 ns) read pulses were employed to fully retrieve the stored photons. However, such pulses risk heating the system, thereby increasing the phonon occupancy of the mechanical oscillator during the retrieval process. To mitigate this effect and maintain the resonator near its mechanical ground state, a reduction in retrieval energy was necessary—albeit at the cost of retrieval efficiency, with only a small fraction of the stored photons being recovered. Specifically, a low-energy read pulse (peak power = –16 dBm, duration = 50 ns) was adopted to minimize thermal excitation during retrieval in subsequent experiments. This pulse configuration was selected based on measurements of the asymmetry in photon count rates by red- and blue-detuned excitations, which indicates a phonon occupancy of 0.17. Since the number of photons scattered under red-sideband excitation is directly proportional to the phonon population, this setting serves as a benchmark for estimating the mechanical occupancy following optical memory.

As shown in Fig. 3(d), weaker read pulses result in the retrieval of fewer photons. The write pulse, comprising control light with a peak power of –32.2 dBm and a duration of 2500 ns, was consistently accompanied by signal photons. These signal photons were generated via an EOM and subsequently attenuated along the optical path, yielding a continuous single-photon detection rate of 0.5 MHz. Following the write pulse, a read pulse was applied. The experiment was conducted at a repetition rate of 5 kHz over the course of one hour to accumulate sufficient statistics. A magnified view of the retrieval region is presented in Fig. 3(d), which shows a clear deviation from the dynamics observed in Fig. 3(c). Only a small number of photons were retrieved. In the retrieval area, the signal peak is enlarged in the inset for clarity.

*Contact author: goubo@swjtu.edu.cn
†Contact author: gwdeng@uestc.edu.cn

To assess whether the phonon occupancy approaches the ground state under various conditions, the photon counts collected using red-sideband read pulses in the absence of any input signal were first used as a baseline reference. Importantly, the power and duration of the read pulses were held constant throughout the measurements to ensure comparability. Subsequently, both red- and blue-detuned pulses, identical in energy and duration, were applied to the device at the same repetition rate. Based on the asymmetry in the number of scattered photons, the phonon occupancy was calculated as $\langle n \rangle = \Gamma_{red}/(\Gamma_{blue} - \Gamma_{red})$, where $\Gamma_{blue}$ and $\Gamma_{red}$ denote photon count rates from blue- and red-sideband read pulses, respectively, integrated over equal time windows. In our experiment, for a read pulse with a peak power of -16 dBm and a duration of 50 ns, the corresponding phonon occupation number was estimated to be $0.171 \pm 0.066$. Since the phonon population is proportional to the anti-Stokes scattering rate, its variation under different experimental conditions can also be inferred from the number of anti-Stokes scattered photons, provided that the read pulse intensity remains constant (See Supplemental Material, Sec. 12). This measurement framework enabled the evaluation of heating induced by the control light pulse and confirmed that the system remains close to the mechanical quantum ground state during the retrieval process.

The retrieval results at different power levels of the write pulses are shown in Fig. 4(a). The intensity of the signal photon pulse is identical to that in Fig. 3(d), with only the power of the control light adjusted. When the number of retrieved photons aligns with the red line in Fig. 4(b), it corresponds to effective phonon occupancy of 0.17. When the control light intensity is -32 dBm, the stored phonon number is 0.32. While the linear nature of the optomechanical interaction does not restrict the number of stored photons, maintaining this sub-unity phonon occupancy ensures that the thermal noise floor is kept below the single-quantum level, a necessary condition for future quantum memory operations. Even accounting for noise from heating, the mechanical resonator is still near the quantum ground state regime [24,36]. Figure 4(b) shows the phonon occupation measurements calibrated via sideband-scattering asymmetry. The dashed arrow highlights the red-sideband scattered photons, which are used as a reference to correlate with the results shown in Fig. 4(a). While increasing the control light power improves storage efficiency, it simultaneously introduces significant thermal noise into the system. These results indicate that the system remains near the mechanical quantum ground state when the control pulse peak power is kept below –30 dBm, despite the presence of control-light-induced thermal excitation. Memory efficiency is defined as the ratio of retrieved photons to stored photons and was calculated accordingly. We chose this internal metric to isolate and evaluate the intrinsic physical limits of the OMIT-based optomechanical interaction. The number of signal photons detected at the SPD was 0.97 per pulse, compared to an input of 41 per pulse, accounting for experimental optical path losses.

Figure 4(c) shows the storage efficiency as a function of control pulse power. Blue diamond markers represent optical pulses of 1050 ns duration that achieve near-complete photon retrieval, while red circular markers correspond to shorter 50 ns pulses used for phonon occupancy measurements with minimal heating effects. Under longer pulse conditions, storage efficiency reaches 4.2%, whereas operation in the quantum ground state limits efficiency to approximately $5 \times 10^{-5}$. Pulse duration critically affects performance: longer pulses reduce spectral broadening, thereby increasing the probability that signals remain within the OMIT transparency window. Figure 4(d) illustrates the relationship between write pulse duration and memory efficiency. It is observed that as the pulse duration increases, storage efficiency improves. For a 2.5 μs optical pulse, the spectral broadening is approximately 0.5 MHz (See Supplemental Material, Sec. 7 [46]), which remains mismatched with the OMIT transparency window bandwidth. Extending the pulse duration further could continue improving the efficiency. However, given constraints in the device's heat dissipation capacity, a write pulse duration of 2.5 μs was chosen. Optimization of write pulse duration and power parameters represents a promising avenue for performance enhancement.

By tuning the control light power from –36.2 to –

*Contact author: goubo@swjtu.edu.cn
†Contact author: gwdeng@uestc.edu.cn

24.2 dBm, a clear trade-off emerges between storage efficiency and thermal noise. Low-energy read pulses mitigate optical absorption heating but reduce storage efficiency to $5 \times 10^{-5}$. The significant difference between the two measured efficiencies is primarily due to the energy limitation of the readout pulse. Therefore, if the heat dissipation issue can be better resolved, higher-energy readout pulses can be employed. Crucially, the spectral alignment between the OMIT window bandwidth and the optical pulse bandwidth plays a pivotal role in determining memory performance. Longer pulses exhibit reduced spectral broadening, which enhances the alignment of the write pulse within the transparency window. This improved bandwidth matching compensates for the reduced transparency depth. While the efficiency of our optomechanical memory currently remains lower than that of atomic ensemble-based EIT protocols, further optimization of device parameters and thermal management strategies is anticipated to substantially enhance memory performance, highlighting the considerable promise of this approach for future scalable and integrated quantum technologies.

## IV. DISCUSSION

In this work, we investigate few-photon memory within one-dimensional optomechanical crystal microcavities, providing evidence for the viability of on-chip absorptive photonic memory via the optomechanically induced transparency effect. Through pulsed light excitation and phonon counting measurements, photon memory near the mechanical ground state was observed, highlighting the potential of solid-state mechanical resonators for quantum information processing. In our optomechanical crystal microcavity, signal photons generated via an EOM sideband are successfully stored and retrieved with an internal efficiency of 4.2%. This operation indicates the platform's compatibility with external photon sources. To accelerate experimental measurements, storage time is set to 150 ns, while future implementations could extend storage times to seconds [26,48] by adding acoustic radiation shield (See Supplemental Material, Sec. 8 [46]) and optimizing the sample. Although current storage efficiency is limited by linewidth matching and thermal dissipation constraints, optimizing pulse parameters and thermal management should further enhance the storage efficiency. Notably, recent advances in 2D optomechanical crystal microcavities [32,33,49-51] have significantly improved heat dissipation capabilities of such devices. Enhanced thermal dissipation and reduced system losses improve the signal-to-noise ratio, minimizing required experimental repetitions and enabling the practical measurement of high-Q resonators with much longer storage lifetimes. The coherence time for this type of nanobeam sample was measured at 34 ± 3.7 μs (See Supplemental Material, Sec. 9 [46]), and can be extended via saturating two-level systems by ac electric field or using alternative materials such as GaP and silicon nitride [52].

For quantum information applications, integrating this memory with external quantum light sources—such as entangled [45] and single-photon sources[43]—is essential. A key challenge is achieving proper bandwidth matching. Fortunately, advances in on-chip light sources are driving improvements in quality factors and brightness[44], thereby producing quantum light sources with bandwidths increasingly aligned with the transparency linewidth. Further bandwidth matching can be accomplished by using a light source built with a free-space optical path [42], which generate sources with narrower bandwidths, or through techniques such as bandwidth compression [53], to bridge the bandwidth gap between the light source and the storage medium. Beyond spectral matching, integrating these heralded sources will natively address our current noise limitations. In practical quantum networks, coincidence detection effectively filters out uncorrelated background noise. Therefore, although our current near-ground-state efficiency is lower than that of mature atomic systems, this coincidence gating ensures that low efficiency primarily limits the event rate rather than degrading the quantum state fidelity. This intrinsic decoupling makes our platform a viable proof-of-principle for an absorptive mechanical interface, paving the way for foundational entanglement distribution studies in the mK regime.

## METHODS

### 1.Device fabrication

The sample is fabricated on a silicon-on-insulator (SOI) substrate consisting of a 220-nm-thick top silicon device layer (<100> orientation), a 3-μm buried oxide layer ($SiO_2$), and a 500-μm silicon handle layer. The device pattern is defined via electron-beam lithography (EBL) and subsequently transferred into the silicon device layer using inductively coupled plasma reactive ion etching (ICP-RIE). To enable optical coupling without imaging—such as in cryogenic experiments—the chip is divided via stealth dicing by laser to expose the waveguide end-facets. Finally, the structure is released by removing the sacrificial oxide layer using a wet etch with hydrofluoric acid (HF), resulting in a suspended optomechanical crystal platform. Prior to low-temperature measurements, an additional cleaning step with piranha solution and a dilute HF dip are performed to reduce surface oxidation.

*Contact author: goubo@swjtu.edu.cn
†Contact author: gwdeng@uestc.edu.cn

**2.Optomechanical sideband signal extraction**

In optomechanical crystal cavities, optomechanical interaction generates optical sidebands. These sidebands, along with the pump laser, produce beat signals detectable by a photodetector, enabling mechanical motion detection. However, when the mechanical oscillator is in its quantum ground state, the sideband signals become extremely faint and require extraction via cascaded optical filters before single-photon detection.

The Fabry–Pérot (F-P) filter, based on multi-beam interference, can be used for such purposes. Here, we employ all-solid-state monolithic etalon fabricated from quartz. Temperature tuning shifts the resonance via refractive index change, allowing the resonant wavelength to be aligned with the optomechanical sideband signal. Two cavity lengths (5.39 mm and 5.86 mm) are used, offering FSRs of 19.5 GHz and 17.5 GHz, respectively, with a linewidth of ~110 MHz and optical extinction about 30 dB. Optical isolators are inserted between stages to prevent back-reflection. Each cavity is temperature-stabilized within 0.01 °C in a copper mount, enabling reliable filtering across the C-band for sensitive sideband detection.

## ACKNOWLEDGMENTS

National Key Research and Development Program of China (grant 2022YFA1405900), National Natural Science Foundation of China (grants 92565107, 62304255, U2441217, 92365112，12574399), Innovation Program for Quantum Science and Technology (grant 2021ZD0302300), Sichuan Science and Technology Program (grant 2024YFHZ0372), and China Postdoctoral Science Foundation (2025M783362).

The authors declare no competing interests.

**Data availability.** The data and relevant computer codes supporting this study are available upon reasonable request from the corresponding author.

*Contact author: goubo@swjtu.edu.cn
†Contact author: gwdeng@uestc.edu.cn

*Contact author: goubo@swjtu.edu.cn

†Contact author: gwdeng@uestc.edu.cn

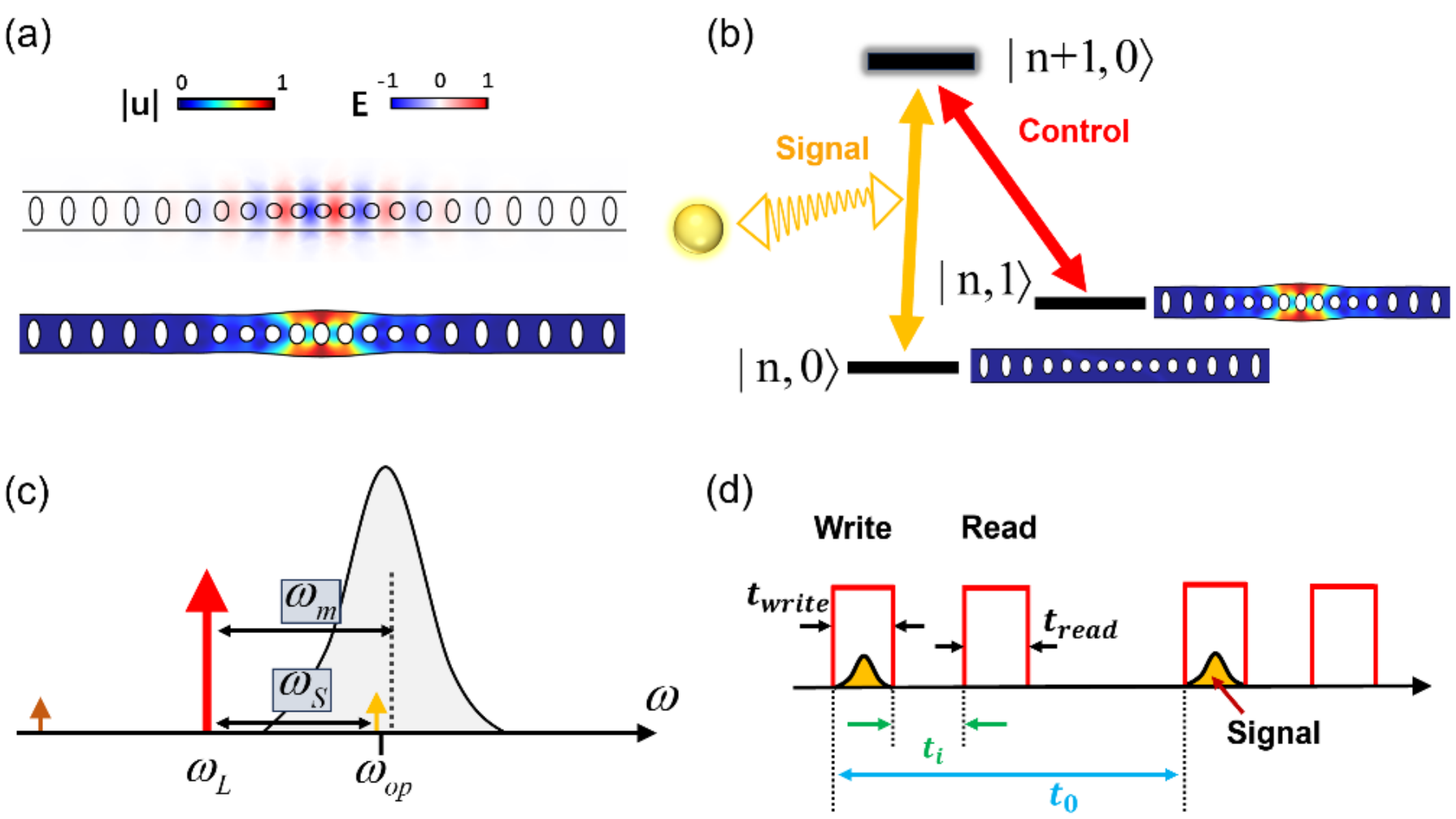

FIG 1. Schematic diagrams of device, OMIT, and the memory process. (a) Optical and mechanical modes of a one-dimensional optomechanical crystal cavity. (b) Energy level diagram for OMIT. Under the influence of a specific optical field, the light absorption of the optomechanical system for the probing light is significantly suppressed, resulting in transparency in originally opaque frequency regions. Here, a red-detuned laser (depicted by the red arrow) can shift the mechanical oscillator's state from |n, 0⟩ to | n, 1⟩, The highest state |n+1, 0⟩ is a metastable state and can also assist in photon absorption (illustrated by the yellow sphere), leading to the transition of the mechanical oscillator's state from |1⟩ to |2⟩. The photon absorption process requires the fulfillment of the two-photon resonance condition, $\Delta\omega = \omega_m$. (c) Illustration of the optical mode near the red-detuned side. A strong control field ($\omega_L$) carries a modulation sideband at the signal frequency $\omega_S$. $\omega_S$ is the modulation frequency of the red-detuned laser, and $\omega_{op}$ is the center of the optical resonance frequency. (d) Time-domain optical signal during the OMIT storage process. During the write phase, both the control pulse and the signal photons are present. During the read phase, only the control field is applied.

*Contact author: goubo@swjtu.edu.cn

†Contact author: gwdeng@uestc.edu.cn

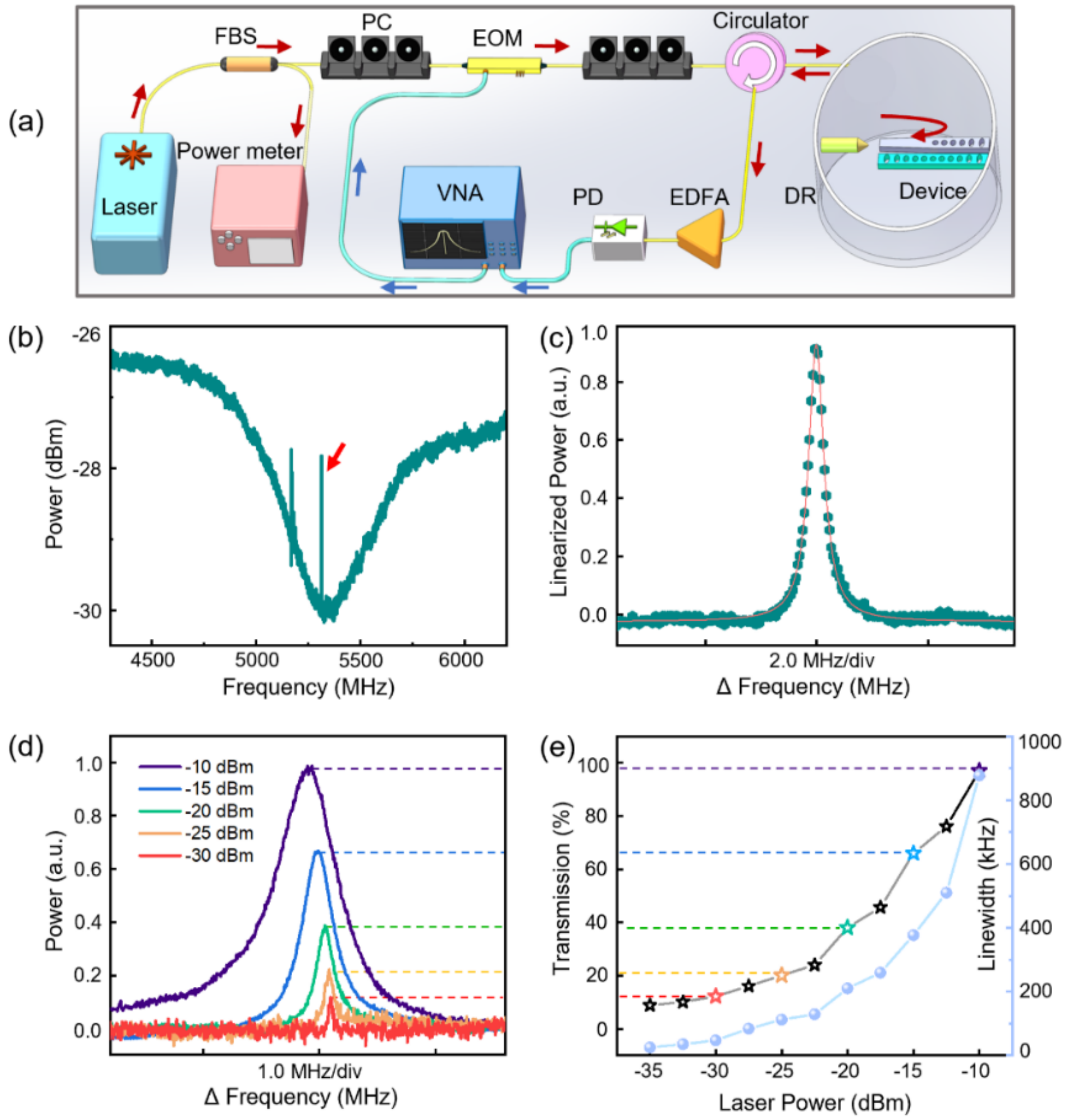

FIG 2. Characterization of the OMIT Phenomenon. (a) Experiment setup. The optical sideband swept across the optical cavity is modulated via one port of a vector network analyzer (VNA) and reflected to a second port for recording the intensity as a function of modulation frequency. EOM, electro-optic modulator. PC, polarization controller; EDFA, erbium-doped fiber amplifier; FBS, fiber beam splitter; PD, photoelectric detector; DR, dilution refrigerator. (b) OMIT transmission spectrum. The dips correspond to optical resonance modes, while the central peak indicates the presence of optomechanically induced transparency. (c) Magnified view of the OMIT peak shown in (b), revealing a transparency window with a width of 332 kHz. (d) Transparency windows measured at various control light powers. (e) Dependence of the transparency window's depth and width on the control light power.

*Contact author: goubo@swjtu.edu.cn
†Contact author: gwdeng@uestc.edu.cn

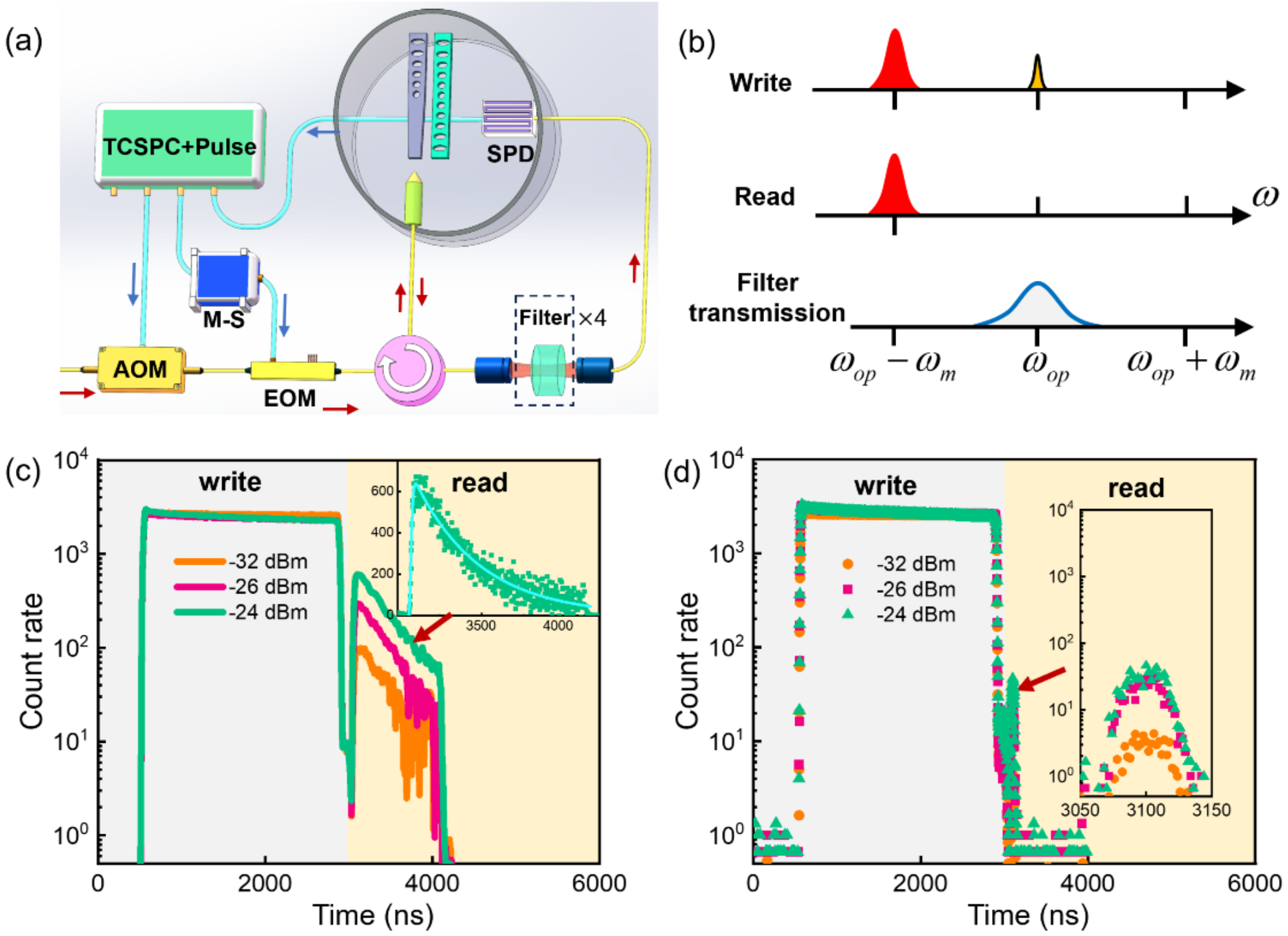

FIG 3. Experimental results of OMIT-based photon memory. (a) Experimental setup. AOM generates pulses, while EOM produces the sidebands of the signal light. The signal and control light pulses are simultaneously injected into the device. The reflected light is filtered and detected by a single-photon detector. AOM: acousto-optic modulator; EOM: electro-optic modulator; SPD: single-photon detector; M-S: microwave source; TCSPC: time-correlated single photon counting. (b) Frequency-domain distribution of the control light, signal photons, and filter bandwidth. (c) Photon storage dynamics, showing both the write and retrieval phases. (d) Retrieval dynamics under reduced control light power. The read pulse has a peak power of -16 dBm and a duration of 50 ns.

*Contact author: goubo@swjtu.edu.cn
†Contact author: gwdeng@uestc.edu.cn

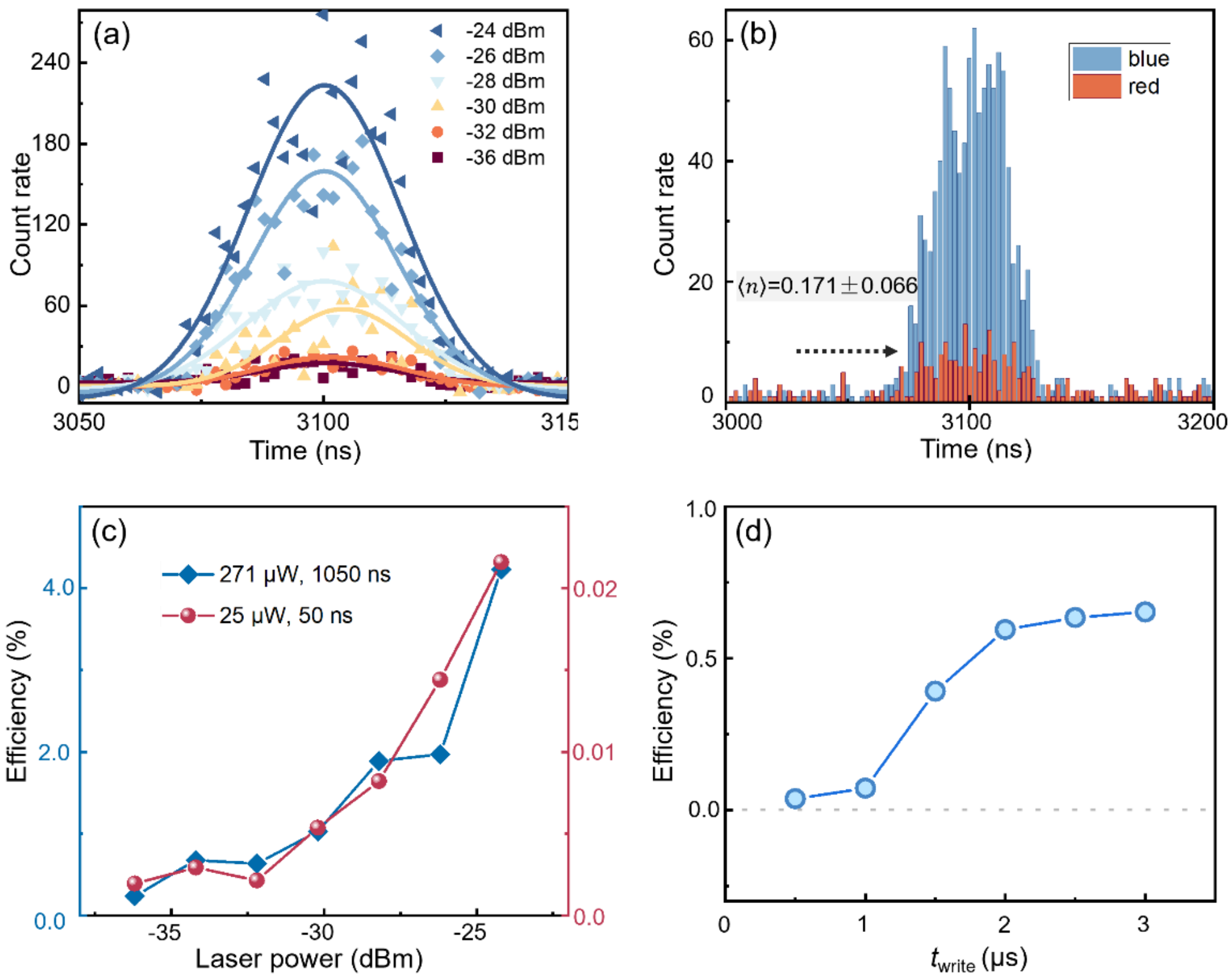

FIG 4. Phonon occupancy and memory efficiency. (a) Retrieved signal traces under different control light powers. (b) Phonon occupancy measured via sideband-scattering asymmetry. The device is driven by blue- and red-detuned optical pulses at frequencies of $\omega_o \pm \omega_m$. Photon count rates, recorded by a single-photon detector, are plotted as a function of the scattered photon's arrival time. (c) Storage efficiency as a function of control light power, measured using two types of retrieval pulses. (d) Storage efficiency as a function of write pulse duration.

*Contact author: goubo@swjtu.edu.cn

†Contact author: gwdeng@uestc.edu.cn